\documentstyle[12pt,aasms]{article}

\slugcomment{{\em Astrophysical Journal}, to appear in 1 September 1995 issue. }

\begin{document}

\title{
Feasibility of Measuring the Cosmological Constant $\Lambda$ and Mass
Density $\Omega$ using
Type Ia Supernovae}

\author{
Ariel Goobar\altaffilmark{1} 
and Saul Perlmutter \altaffilmark{2}
}
\altaffiltext{1}{Physics Department, Stockholm University, Box 6730, 
S-113 85 Stockholm, Sweden; ariel@physto.se}
\altaffiltext{2}{Center for Particle Astrophysics, Berkeley and
 Lawrence Berkeley Laboratory, 50-232, 
Berkeley, CA  94720; saul@lbl.gov}




\begin{abstract}
We explore the feasibility of resurrecting the
apparent magnitude-redshift
relation for a ``standard candle''  to
measure the cosmological constant and mass density.
We show that type Ia supernovae, if measured with 0.15 mag uncertainty
out to a redshift of $z=1$, may provide a good standard candle
or calibrated candle for this
purpose.
The recent discovery of probable type Ia supernovae 
in the redshift range $z=0.3$ to
0.5 (Perlmutter {\em et al.} 1994a,
and 1994b) indicates that
the flux of optical photons
from these events can be measured this accurately.
The 7 distant supernovae discovered to date
do not by themselves distinguish between different
cosmological models, however the further
discovery of about 50 type Ia supernovae at redshifts
in the range $0.5 \leq z \leq 1.0 $ could strongly constrain the allowed range
of these parameters. We estimate that the
follow-up photometry necessary
for this measurement would be on the order of $20 - 70$ hours
of time on a 10-meter class telescope at a site with good seeing.
\end{abstract}

\keywords{cosmology: observations, cosmological constant---supernovae}

%
%

\section{Introduction}

\vspace{-0.18in}
Recent attention to the problem of measuring or
bounding the cosmological
constant $\Lambda$ has yielded inconclusive results.
The review article of Carroll, Press, \& Turner (1992)
surveyed the observational status of the
cosmological constant based on (1) the existence of high-redshift objects,
(2) the ages of globular clusters and cosmic nuclear chronometry, (3) galaxy
counts as a function of redshift or apparent magnitude, (4) dynamical tests
(clustering and structure formation), (5) quasar absorption line statistics,
(6) gravitational lensing counts and statistics, and (7) the astrophysics
of distant objects. The conclusion of this exhaustive survey was
that the current best ``observationally secure''
bounds on the cosmological constant are $-7 < \Lambda /(3 H_0^2) < 2$,
leaving a wide range of possible cosmological models to choose from.
In fact, we still do not know if we live
in an infinite universe that will expand infinitely
or a finite universe that at some point will halt its expansion and
recollapse.
In this paper we explore the feasibility of resurrecting the
apparent magnitude-redshift
relation for a ``standard candle'' as an eighth method to add to this
arsenal of measurement techniques.

\vspace{0.04in}
The early work on the implications of cosmological models on the apparent
magnitude-redshift ($m$-$z$) relation of
a standard candle, the first-ranked cluster galaxies,
did consider the possibility
of a non-zero cosmological constant (e.g.,
Solheim 1966,
Stabell \& Refsdal 1966).
As the difficulties
of studying evolutionary effects for these galaxies became clear, the
range of cosmological models considered narrowed to just those with a vanishing
cosmological constant (e.g., Peach 1970).
The equations of galaxy evolution and the
deceleration parameter $q_0$ (or equivalently the mass density of the
universe $\Omega_M$) were considered complications enough in these
$m$-$z$ studies.  The most recent work has generally been considered to be
more a study of evolution than a measurement of $q_0$ or $\Omega_M$ (for
a review see Sandage 1988).
In the past few years
new evidence has been put forward suggesting that a group of
type Ia supernovae (SNe Ia) can be identified that are excellent
standard candles or calibrated
candles.  There is reason to believe that evolution effects should
be much less significant for SNe Ia than for first-ranked cluster galaxies and
that even if present such effects may be distinguishable
on an event by event basis.  The past few years have also seen the
start of searches for distant SNe, resulting in the discovery and study
of 7 SNe Ia at redshifts in the range $z=0.3$ to 0.5
(Norgaard-Nielsen {\em et al.} 1989; Perlmutter {\em et al.} 1994a, 1994b).
This is clearly an opportune time
to reconsider the
use of standard candles to measure $\Lambda$.

In this paper we first review the current understanding of the usefulness
of a sub-group of SNe Ia as standard candles, and the possibility of
further ``calibrating'' these candles using lightcurve decay-time or shape.
We then discuss the use of standard candles to measure the cosmological
constant and mass density.
Some of the earliest papers that
treated the $\Lambda \neq
0$ case pointed out
that the magnitude--redshift
measurement was insensitive to $q_0$ at certain redshifts while still
sensitive to $\Omega_M$ (e.g. Refsdal, Stabell, \&  Lange 1967).
We propose to take advantage of this redshift dependence
to measure $\Omega_M$ and $\Lambda$ simultaneously.
The special case of a ``flat'' universe, as
implied by the inflationary theories
of the universe, is discussed separately.  We then draw
conclusions about the observational
requirements and hence the
feasibility of a new measurement of $\Lambda$ and $\Omega_M$ using
SNe Ia.

\section{Type Ia supernovae}

\vspace{-0.27in}
There is much evidence indicating that a distinguishable majority of
type Ia supernovae are likely to be good standard
candles.
The problem of estimating the intrinsic dispersion of SNe Ia, however, has been
clouded by the inclusion of supernovae with peculiar
spectra or lightcurves, supernovae showing clear evidence
of host-galaxy extinction, and supernovae that had very large uncertainties
on their photometry measurements.
For a subsample of well-measured ``local'' 
SNe Ia that do not have peculiar
spectra or lightcurves and do not show clear evidence of extinction,
the {\it observed} dispersion is
$\sigma_V = \sigma_B = 0.3 $ magnitude in both the $V$ and $B$ bands (Vaughan
{\em et al.} 1994).
This dispersion of these ``normal'' SNe Ia
is completely accounted for by measurement errors
(most of this scatter is probably due to the relative-distance measurement
error) and thus the {\it intrinsic} dispersion is likely to be
smaller than this. Sandage \& Tammann (1993) use Monte Carlo modeling of
Malmquist bias to argue
that the largest intrinsic dispersion
for type Ia supernovae which is compatible
with the observed selection effects for nearby supernovae is
$\sigma^{intrinsic}_{M_V}   \approx 0.2 $ mag.

Vaughan {\em et al.} proposed that their criteria for ``normal'' SNe Ia
be tested on an independent set of SNe to confirm the small observed
observed dispersions $\sigma_V$ and $\sigma_B$.  
Hamuy {\em et al.} (1994) presented such an independent,
new set of SNe Ia, including both ``normals'' and ``peculiars,''
with smaller measurement
errors.  In particular, the relative-distance measurement error was smaller,
because this set of SNe was discovered at redshifts $z \approx 0.01$ to 0.1 
where the peculiar velocities are negligible with respect to the Hubble 
flow.  Selecting just the ``normal'' SNe Ia from this set, using
the criteria of Vaughan {\em et al}, results in an even narrower observed
dispersion of $\sigma_V = 0.23$ mag in the $V$ band and 
$\sigma_B = 0.25$ mag in the $B$ band (Vaughan, Branch, \& Perlmutter
1994).

Hamuy {\em et al.} (1994) and Riess, Press, \& Kirshner (1994) also reported a 
correlation of lightcurve decay-time or lightcurve shape with peak absolute
magnitude for this set of SNe Ia.   (Note that this correlation would not be easily found
in the earlier set of SNe Ia with larger measurement errors, although
Phillips 1993 did report such a relation for a small sample of 
well-measured peculiar and normal SNe Ia.)
Using this correlation to provide a ``calibration'' of the
SN Ia  standard candle may make it possible to include peculiar SNe Ia in
distance measurements.  The correlation also appears to hold within the 
``normal'' SN Ia subset,  allowing even this subset's 
already narrow dispersion to be further reduced after calibration,
yielding $\sigma_V$ as low as 0.12 (Hamuy {\em et al.} 1994) or 0.21 (Riess,
Press, \& Kirshner 1994).  Although this calibrated $\sigma_V$ 
would imply a still
smaller intrinsic dispersion, for this paper we will take the intrinsic
dispersion to be the ``uncalibrated'' value, which is bounded by the
observed dispersions to be
$\sigma^{intrinsic}  < 0.25$ in the $V$ or the $B$ band.  This is 
a conservative value, given that the 
{\em observed} dispersions quoted in Vaughan, Branch, \& Perlmutter
(1994) are less than or equal to this.

If SNe Ia are to be more useful as cosmological standard candles than the
first-ranked cluster galaxies have been, they either must not evolve
in absolute magnitude or this evolution must be easily detected and
characterized.  There are at least two reasons suggesting that SN Ia
standard candles should not founder on the evolution problem:

(1) Unlike first-ranked cluster galaxies, SNe Ia are dynamic events that
display their internal composition and physical state through the many
spectral lines that appear, shift in velocity, and disappear, and also
through the photometric lightcurves in various wavelength bands.  It is
possible to observe each individual SN Ia, match its spectra over time and
lightcurves against those of nearby SNe Ia, and check for subtle changes
from the range of normal SNe Ia.  These changes are very likely to be
{\em more} sensitive to
the details of the precursor star and environment than the peak absolute
magnitude is, and thus can provide ``early warning''
before there are differences
large enough to affect the absolute magnitude significantly.
For example , the lightcurve decay-time or shape and the spectral absorption
line velocities both appear to be sensitive indicators of explosion
strength.

(2) SNe Ia have been discovered in a wide range of nearby galaxy types.  This
variation in host galaxy environment can be used as a surrogate for the
variation that would be expected due to evolution.  This has been done, for
example, by Branch \& van den Bergh (1993), who suggest that Si II absorption
line velocity may be
correlated with host galaxy type.  Branch \& van den Bergh
did not see a correlation with absolute magnitude in this case, but such
studies of nearby supernovae
can in principle detect, and 
provide tests for, evolution of absolute magnitudes.  Ideally these tests would
make it possible to distinguish degrees of 
evolution on a supernova-by-supernova basis.

Even if the SNe Ia themselves do not evolve, it is possible that the host
galaxy dust may evolve, thus changing the apparent magnitude with redshift.
Although very careful color photometry should provide checks for this
effect, it is probably easier to  compare SNe Ia in different galaxy types
(both nearby and distant), once again using these types as surrogates, this
time for evolution of host galaxy dust.  So far there does not appear to be
such an effect for a range of nearby galaxy types.

These evolution tests will provide the underlying proof of SNe Ia as
standard candles or calibrated candles, 
and could of course someday find some SNe Ia exhibiting
evolution effects that cannot be easily corrected.  It is important to
re-emphasize, however, that SNe Ia are unusual standard candles in having
such tests available on an individual basis: each SN Ia can be accepted or
rejected by itself.

\section{Constraining the parameters by standard candle luminosity distance}

\vspace{-0.27in}
For an object of known absolute magnitude $M$, a measurement of
apparent magnitude $m$ at a given redshift is sensitive to the universal
parameters $\Omega_M$ and $\Omega_\Lambda \equiv \Lambda /(3 H_0^2)$
through the
luminosity distance $D_L$:
\begin{equation}
  m = M + 5 \log[{D_L(z;\Omega_M,\Omega_\Lambda)}] + K + 25,
\end{equation}
where the $K$-correction in the equation appears because the emitted and
detected photons from the receding object have different wavelengths.
The dependence of $D_L$ on $\Omega_M$ is
different from the dependence on $\Omega_\Lambda$, entering with different
powers of $z$:
\begin{equation}
D_L(z;\Omega_M,\Omega_\Lambda) =
\frac{(1 + z)}{H_0 \sqrt{|\kappa| }} \; \; \; {\cal S}\! \left (
  \sqrt{|{\kappa}| } \int_0^{z} \left [(1+z^\prime)^2(1+\Omega_M z^\prime)-
   z^\prime (2+z^\prime ) \Omega_\Lambda \right]^{-\frac{1}{2}} dz^\prime
  \right ),
\end{equation}
where, for $\Omega_M + \Omega_\Lambda < 1$,
${\cal S}(x)$ is defined as $\sin(x)$ and $\kappa
= 1 - \Omega_M - \Omega_\Lambda $;
for $\Omega_M + \Omega_\Lambda > 1$, ${\cal S}(x) =
{\rm sinh}(x)$ and $\kappa$ as above; and for $\Omega_M + \Omega_\Lambda = 1$,
${\cal S}(x) = x$ and
$\kappa =1$.

\begin{figure}[tb]
\plotone{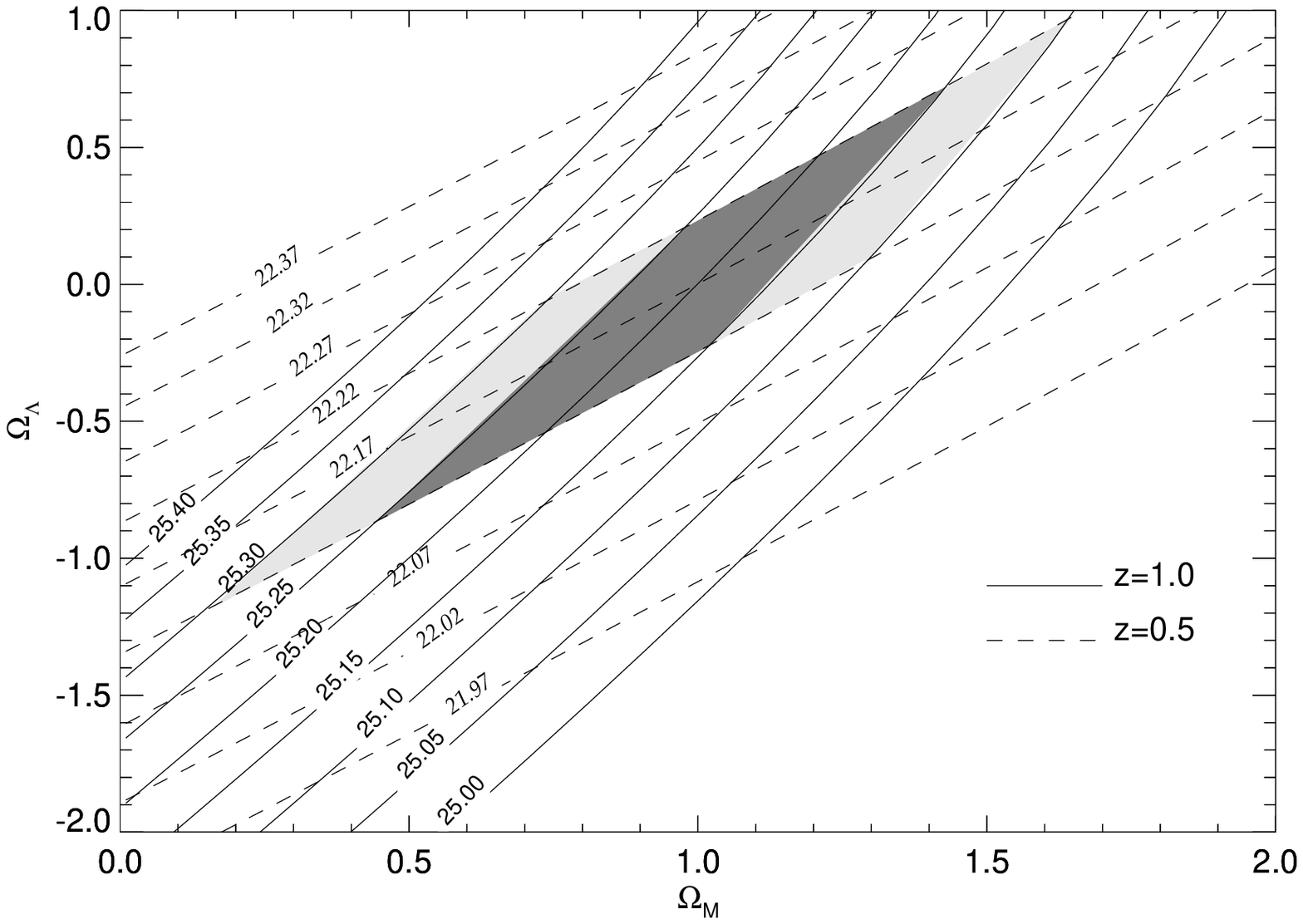}
 \caption[lvsm]{ \small Contours of constant apparent magnitude (R-band)
predicted for an example standard candle with absolute magnitude (B-band)
$M_{B} = -18.86 +5 \log(H_0/75)$.  The dashed lines show the predicted apparent
magnitude, including $K$-corrections, 
 for a standard candle at $z=0.5$ and the dashed lines are for
$z=1$.   
The dark shaded region shows the ``allowed'' region of 
$\Omega_\Lambda$-versus-$\Omega_M$ parameter space if an apparent magnitude
of $m_R=22.17 \pm 0.05$ were measured at $z=0.5$ and $m_R=25.20 \pm 0.05$ were
measured at $z=1$. Adding the faint shaded region implies a 0.1 magnitude
uncertainty for supernovae at $z=1$.}
 \label{contours}
\end{figure}

Using Equations (1) and (2) we can predict the apparent magnitude of a
standard candle measured at a given redshift for any pair of values of
$\Omega_M$ and $\Omega_\Lambda$.
Note that the value of the Hubble parameter drops out of the equations as
it appears both in the expression for the luminosity distance and
in the determination of the absolute magnitude of the standard candle based
on nearby apparent magnitude--redshift measurements.
Figure \ref{contours} shows the contours of
constant apparent magnitude in  the R-band
on the $\Omega_\Lambda$-versus-$\Omega_M$ plane,
for the cases of $z=0.5$ and $z=1$, where
we have taken
the absolute luminosity of type Ia supernovae to be $M_B = -18.86 \pm 0.06
+ 5 \log{(H_0/75)} $ (Branch \& Miller 1993, Vaughan {\em et al.} 1994).

When an actual apparent magnitude
measurement is made of a standard candle, for example at $z=0.5$,
the range of possible values of $\Omega_M$ and
$\Omega_\Lambda$ are narrowed to a single contour line on Figure \ref{contours}
(dashed lines for
$z=0.5$).
Given some uncertainty in the apparent magnitude measurement,
the allowed range of $\Omega_M$ and $\Omega_\Lambda$ is
given by a strip between two
contour lines.
Two such measurements for standard candles at different redshifts (for
example $z=0.5$ and $z=1$) can
define two strips that cross in a more narrowly constrained ``allowed'' region,
shown as a shaded rhombus in Figure \ref{contours}.
The darker shaded region in
the plot corresponds  to the result of
measurements with 0.05 mag uncertainty in a
flat universe with vanishing cosmological constant, while
the faint region allows for a 0.10 mag uncertainty at $z=1$. 
Note that the one-standard-deviation error region is limited by an
ellipse rather than the rhombus in Figure \ref{contours}. To 
simplify this figure and the following two figures, we have not
drawn the 1$\sigma$ error ellipse. 

\begin{figure}[tb]
\plotone{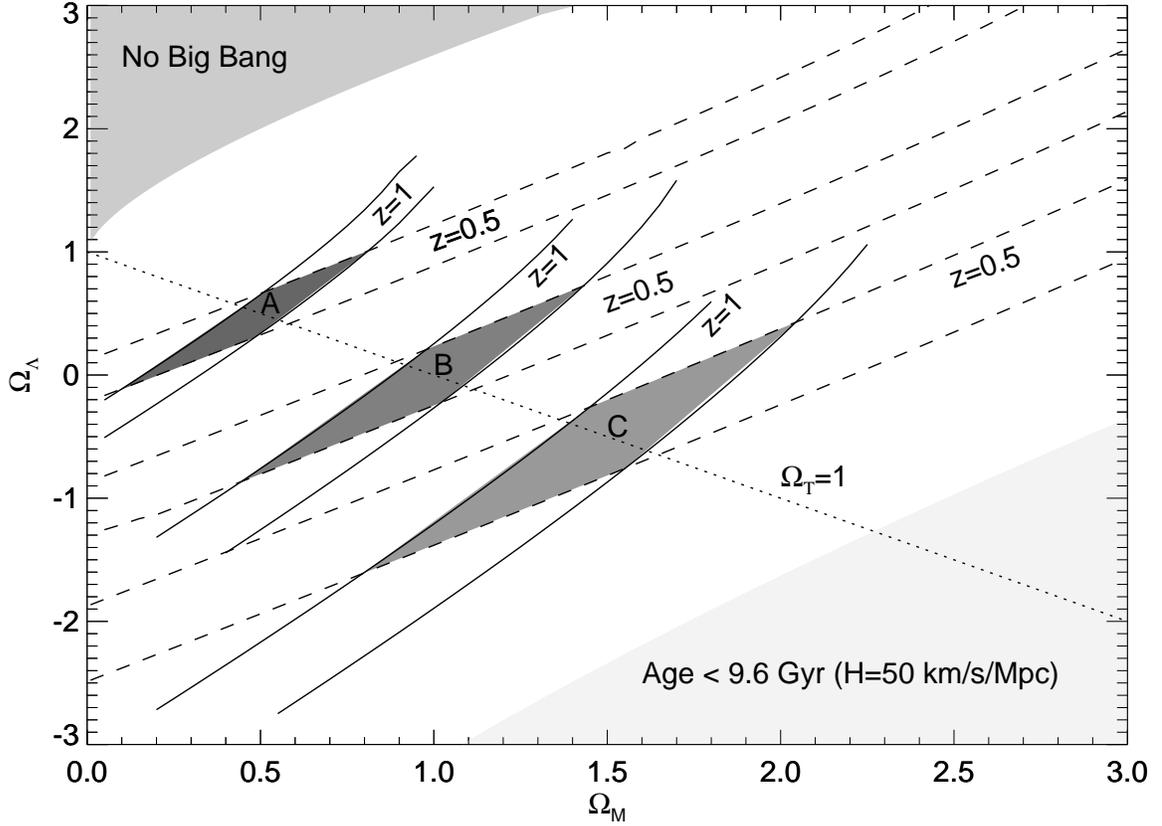}
 \caption[lvsm]{ \small The map of parameter space
 for $\Omega_\Lambda$ and  $\Omega_M$. The top and bottom shaded areas
 are ruled out by observations (see text). The solid lines show the
 enclosed band that 
 a 0.05 mag measurement a standard candle at $z=1$ would imply for
 three different universes. Similarly, the dashed lines correspond to
 the same standard candle at $z=0.5$. The regions A, B and C give
 the allowed parameter space 
for the cases when the parameters
are ($\Omega_\Lambda$=0.5, $\Omega_M$=0.5) for A, 
($\Omega_\Lambda$=0.0, $\Omega_M$=1.0) for B and
($\Omega_\Lambda$=-0.5, $\Omega_M$=1.5) for C.
}
 \label{general}
\end{figure}

\begin{figure}[tb]
\plotone{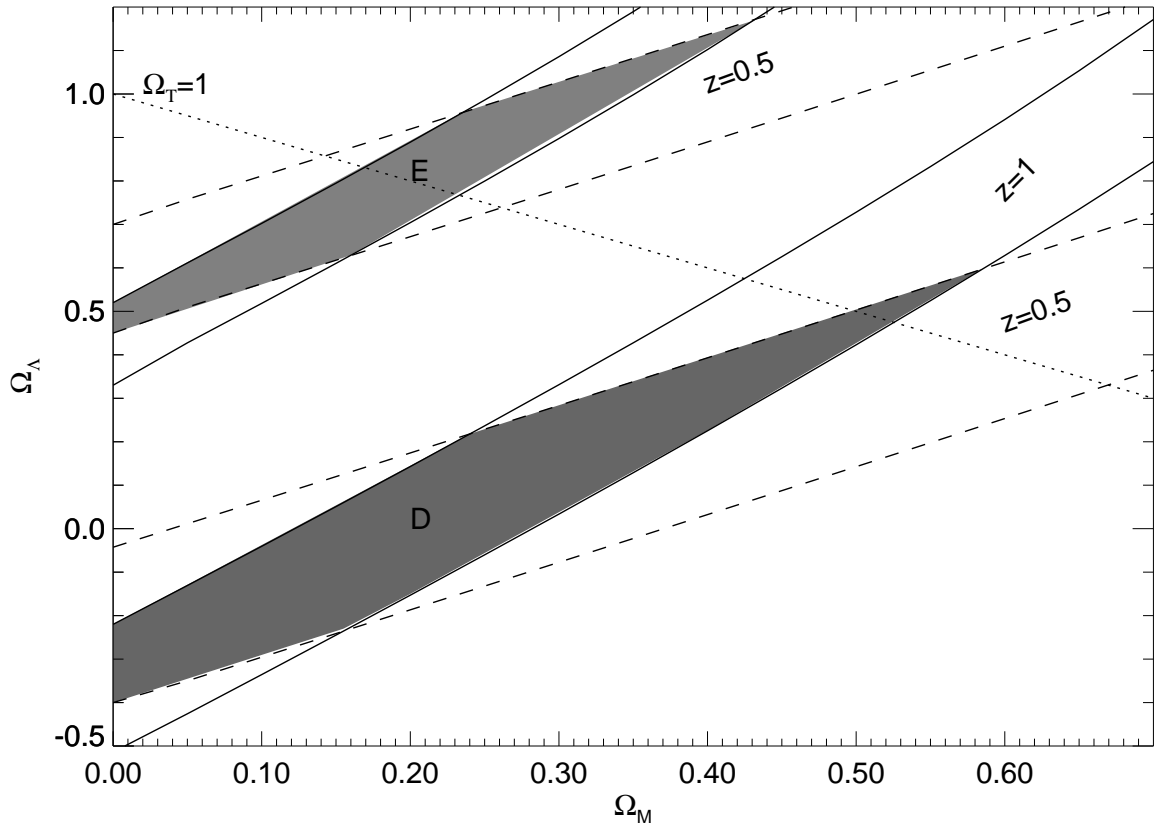}
 \caption[lvsm]{ \small The map of allowed parameter space
 for $\Omega_\Lambda$ and  $\Omega_M$. The region D corresponds to
 $\Omega_\Lambda$=0 and $\Omega_M$=0.2 . E corresponds to $\Omega_\Lambda$=0.8
 and $\Omega_M$=0.2 }
 \label{empty}
\end{figure}

In the case where a standard candle is measured at $z=0.5$ and $z=1$,
Figure \ref{general} shows the allowed regions
for $\Omega_\Lambda$ and $\Omega_M$ for a set of three example universes
{\it superposed} on the same graph (i.e. the
actual measurements would result in
only one of the shaded regions A, B, or C).
Note that on this graph, very large positive
values of $\Omega_\Lambda$ are ruled out because they would imply
a ``bouncing'' (no Big Bang) universe, as discussed in
Carroll, Press, \& Turner (1992).
Also extremely large values
of $\Omega_M$ combined with negative $\Omega_\Lambda$ are ruled out
because they would imply a universe younger than
the oldest heavy elements, which have been dated to be 9.6 Gyr
(Schramm 1990).
The shaded regions
correspond to
hypothetical results in a universe with parameters
($\Omega_\Lambda$=0.5, $\Omega_M$=0.5) for A,
($\Omega_\Lambda$=0.0, $\Omega_M$=1.0) for B, and
($\Omega_\Lambda$=-0.5, $\Omega_M$=1.5) for C. These examples all
correspond to flat
universes, but with different contributions from matter and cosmological
constant.
Similarly, figure \ref{empty} shows how this method would distinguish
the case D ($\Omega_\Lambda$=0.0, $\Omega_M$=0.2) from E
($\Omega_\Lambda$=0.8, $\Omega_M$=0.2).

In practice, more than two apparent magnitude measurements at two
redshifts would be used for this measurement.  A global fit of Equations
(1) and (2) to the measurements would then yield best-fit contours on the
$\Omega_\Lambda$-versus-$\Omega_M$ plane.  
Figures \ref{contours} through
\ref{empty}, however, give a direct understanding of how good the
measurement errors need be to constrain $\Omega_\Lambda$ and $\Omega_M$:
the accuracy of
the magnitude measurements translates into a region in the $\Omega_\Lambda$
versus $\Omega_M$ parameter space
approximately as $(\Delta \Omega_\Lambda \times \Delta \Omega_M) \propto
(\sigma^{z=0.5}_m \times \sigma^{z=1.0}_m)$.
Note that these magnitude errors are
the combined error of the apparent magnitude
measurement at redshift $z$, the absolute magnitude estimate for
the standard candle used, and the intrinsic dispersion of SNe Ia.
We see from the figures that a combined measurement
error of $\sigma_m \le 0.05$ mag  significantly constrains $\Omega_\Lambda$ and  $\Omega_M$.

In this paper we assume that
the photometric measurements are going to be sufficiently precise
that the intrinsic
dispersion of SNe Ia dominates, $\sigma^{intrinsic} < 0.25 $ mag
(in section 5 we discuss the observational requirements to achieve 
this photometric accuracy).
In order to make the $\pm$0.05 mag measurement at $z=0.5$ and $z=1$ shown in figures 
1 through 3, 
we thus must have a sample of at least 25 supernovae at each redshift.

\section{$\Omega_\Lambda$ in the flat universe case}

\begin{figure}[tb]
\plotone{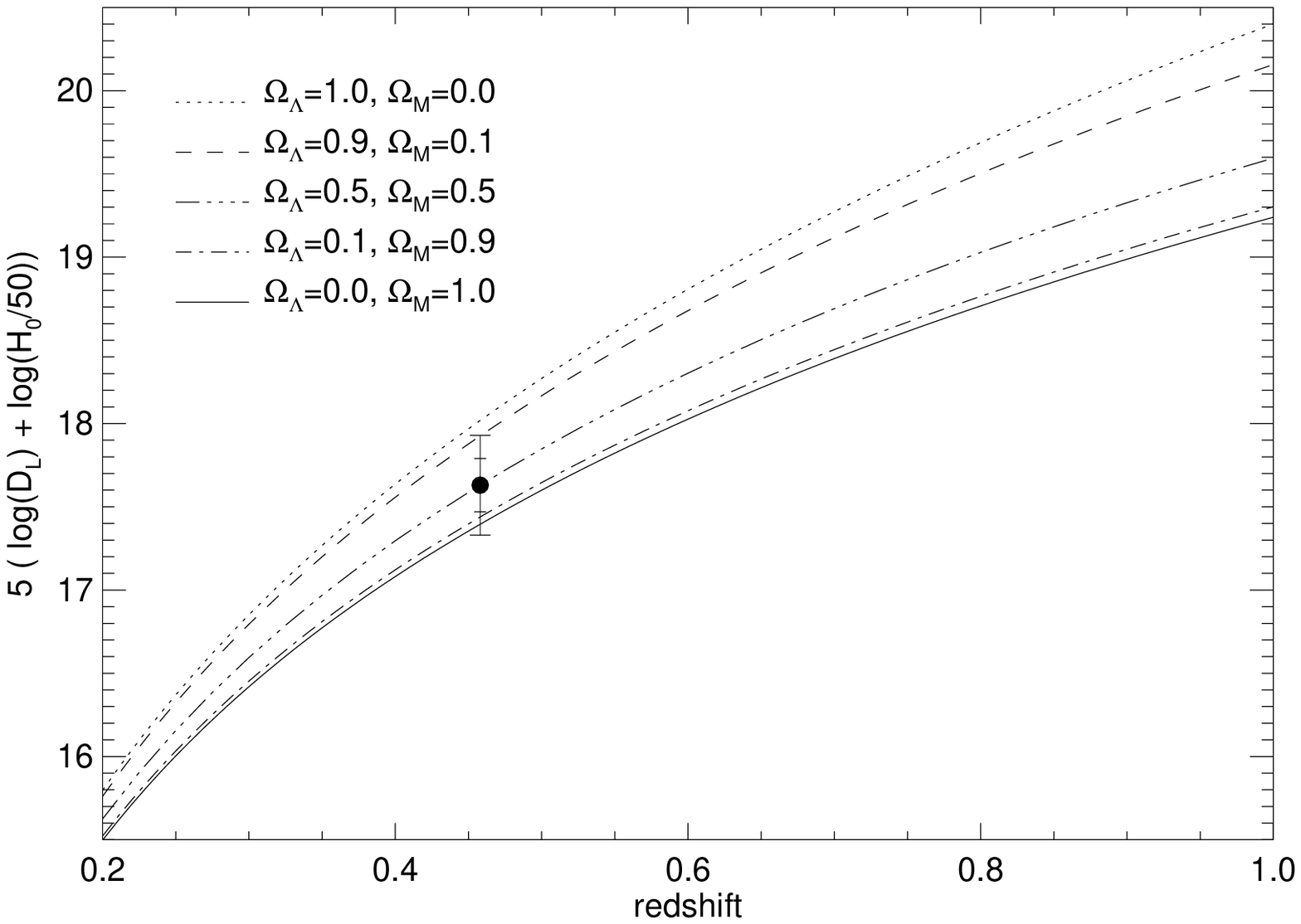}
 \caption[dlvsz]{\small Luminosity distance as a function 
  of redshift for various values of $\Omega_M$ and $\Omega_\Lambda$ in
a flat universe ($\Omega_M + \Omega_\Lambda = 1$). The
  filled circle corresponds to $(m - M - K - 25)$
  for SN1992bi (Perlmutter {\em et al.} 1994a),
where the smaller error bar is due
  to the photometry measurement error, $\sigma^{photometry} \approx 0.15$,
 and the larger 
error bar includes a 0.25 magnitudes 
  intrinsic dispersion for type Ia SNe.}
 \label{magvsz}
\end{figure}

\vspace{-0.27in}
An important special case to consider is the ``flat'' universe predicted by 
the inflationary theories, where the
total energy density of the universe 
$\Omega_T \equiv \Omega_\Lambda
+ \Omega_M = 1 $.  [The other special case with $\Lambda = 0$ has been 
discussed in Perlmutter {\em et al.} (1994a).]
In a flat universe, the apparent magnitude of a standard candle
as a function of redshift is extremely sensitive to 
$\Omega_\Lambda$. Figure	
\ref{magvsz} shows the theoretical curves for the luminosity distance
as a function of redshift for flat universes.
A measurement of the apparent magnitude of a standard candle
at $z=1$ would strongly constrain the
cosmological constant and thus test inflationary models. 
As an example, for the case in which $\Omega_T $ is dominated by 
$\Omega_\Lambda$,  
it could be measured with $\sim$10\% accuracy even with $\sigma_m$
as large as 0.25 mag. 
The ratio of photon flux 
for the $\Omega_M$-dominated versus the $\Omega_\Lambda$-dominated case is
about a factor of three for a standard candle at $z=1$. 

At redshift $z=0.458$, where the most distant type Ia supernova was 
found, the total measurement error, $\sigma_m \approx 0.3$ mag (including
the uncertainty in the photometry, $\sigma^{photometry} \approx 0.15$ 
mag, as well as the uncertainties in the $K$-correction and the intrinsic  
dispersion of type Ia SNe), yielded a 
1$\sigma$ allowed interval of
 $ -0.2 < \Omega_\Lambda < 0.9 $ for  $\Omega_T =1$.  This allowed interval is
shown by the data point and outer error bar
in Figure \ref{magvsz}. Note that this particular supernova
did not have the color measurements that would make it possible to
distinguish host galaxy extinction or
a peculiar supernova, and therefore this provides only a  
demonstration data point. 

\section{Observing requirements}

\vspace{-0.27in}
The analysis of the photometry of SN1992bi showed that one can
measure the apparent $R$ magnitude at peak 
of a supernova at $z = 0.458$ 
with a photometric uncertainty $\sigma^{photometry} \approx 0.15$ mag 
(Perlmutter {\em et al.} 1994a).
Using a 2.5 meter telescope and a ``thick'' CCD (peak quantum 
efficiency $\sim$43\% at 650 nm), a total of 135 minutes of exposures
were required, 90 minutes distributed over 
the four months near peak and a reference image of
45 minutes one year after peak. The average seeing was 
approximately 1.5 arcsec. For SN1992bi, the uncertainty
at peak relative to the reference image was only 0.06 mag, and the
error on the reference image photometry of the host galaxy 
dominated. Clearly, the longest
single exposure should be the one of the reference image of the 
host galaxy after the SN has faded.  In order to take advantage of the
further magnitude calibration from lightcurve decay-time or lightcurve shape, 
this series of
observations must begin before maximum light; the search technique of
Perlmutter {\em et al.} (1994a, 1994b) makes this possible on a 
systematic basis.

Scaling to a 10-meter class telescope, at a site such as Mauna Kea 
with 0.75 arcsec median seeing and with a thinned CCD,
we estimate that the uncertainty in apparent magnitude of distant
supernovae at $z=0.5$ ($\sim$0.2 mag fainter) can be kept below $\sigma^{photometry} = 0.15$ 
magnitudes with 
$ 1.5 \epsilon $ minutes of photometric
measurements, where $\epsilon$ accounts for the scaling
factors:
$ \epsilon =  
 ({\rm seeing}/{0.75"})^2 \;\;
 ({10{\rm m}}/{\rm aperture})^2. $
The photometric
uncertainty is dominated by the sky background at these high redshifts, 
typically more than 4 magnitudes brighter than the counting rate from the
SN and the host galaxy.  Mauna Kea and La Palma, where SN 1992bi was observed,
have essentially the same sky brightness, but at a different site exposure
time would scale with sky, too, as $10^{0.8 (sky_1 - sky_2)}$.

Observing $N=25$ supernovae at $z = 0.5$ would
require less than 1$\epsilon$ hour of 10-meter telescope 
photometry time.
The overall measurement uncertainty would then be
$ \sigma_m = N^{-1/2} \;\;
[ (\sigma^{intrinsic})^2 + (\sigma^{photometry})^2 ]^{1/2} \;\; \le 
0.05$ mag,  for $\sigma^{intrinsic}<0.25$ mag, and neglecting the much smaller 
error in the mean SN Ia absolute magnitude. This is the value of
$ \sigma_m$ discussed in Section 3 and shown in the dark-shaded regions of Figures 1 to 3.

For a type Ia supernova at $z=1$, $5 \log{ D_L } $ is
about 2 magnitudes fainter than for $z=0.5$ (see Figure \ref{magvsz} for 
the effect of different cosmologies on this distance modulus).
Although the choice of the 
$R$ filter is well suited for the $z=0.5$ supernovae, the $I$ filter
is more appropriate for $ z > 0.85$, 
because the rest-frame flux from type Ia supernovae
falls rather steeply below $\sim$300 nm. 
The sky is approximately 0.8 magnitudes brighter in the $I$-band,
but the difference of zeropoints between
the $I$ and $R$ band is $-0.8$ magnitudes, so there are roughly the same
number of sky background {\em photons per second}
in both $R$ and $I$ in spite of the
difference in magnitudes (Massey {et al.} 1995).
Taking into account a reduction of the quantum efficiency
by a factor of $\sim$2 above 800 nm, it would take approximately  
2$\epsilon$ hours of observing time per supernova at $z=1$
to obtain $\sigma^{photometry} \le 0.15$ mag 
uncertainty in the apparent magnitude.

To achieve an overall measurement uncertainty
of $ \sigma_m \le 0.05$ mag would then require 34 SN Ia, or
$\sim$70$\epsilon$ hours
of 10-meter photometry
time.  Alternatively, $ \sigma_m \le 0.1$ mag could be achieved
with only 9 SN Ia observed at $z=1$, requiring $\sim$ 18 $\epsilon$ observing
hours.
Note that a $\sigma_m = 0.1$ mag uncertainty at $z=1$ still 
yields quite useful bounds on the $\Omega_M$ versus $\Omega_\Lambda$ plane 
as shown by the faint-shaded region of Figure \ref{contours}.

The time needed in order 
to {\it find} tens of supernovae is significantly larger.  For example,
at a 10-meter
telescope about 15$\epsilon$ minutes would be needed to find a supernova
at $z \approx 0.5$, using a wide-field camera such as the four-CCD
mosaics currently being commissioned at several observatories. 
Using the 2.4-meter Hubble Space Telescope 
as suggested by Colgate (1979)  to study
high-redshift SN would not significantly 
diminish the length of exposures needed for SN at redshifts $z \lesssim 1$
(see Nelson, Mast, \& Faber 1985 for Keck-HST comparisons).

Based on a 1-hour spectrum of a supernova at $z= 0.425$ 
observed at a 3.6-meter telescope (Perlmutter {\em et al.} 1994b), 
we estimate that 10 hours of 10-meter telescope time are required 
to obtain a spectrum of a supernova at $z=1$, and
only 15 minutes for
supernovae at $z=0.5$. 
For the first set of high redshift SNe, these spectra 
would be necessary in addition to color photometry to check
identification and evolution.  If these spectra show no surprises, it
may be possible to spot check the subsequent SN spectra and use multicolor
lightcurves instead.

In this estimate of the observation time required, we have implicitly included
the $K$-correction by moving to a longer-wavelength band for the higher
redshift measurements.  An important calibration step in the actual
experimental protocol for this measurement will be the careful determination of
the $K$-correction for each SN studied.  Currently available spectra of nearby
SNe allow traditional $K$-correction estimates 
(e.g., corrections for light emitted in the $B$ band at high redshifts
to the light observed
in the $B$ band) to be made with reasonable accuracy ($<0.05$
mag) out to redshifts of order $z \approx  0.2$, within less than 20 days 
(SN rest frame) of maximum
light (Hamuy et al 1993).  A generalization of the $K$-correction that
corrects for light emitted in the $B$ band, for example, at high
redshifts, but observed in the $R$ band can be calculated with this same
accuracy for objects out to at least 
$z = 0.6$ (Kim, Goobar, \& Perlmutter 1995).  
However, for the most accurate
corrections, particularly at high redshifts, it will be important to make
further well-calibrated observations of a number of newly discovered  nearby
SN Ia spectra and lightcurves, to ensure that any supernova-to-supernova
differences are sampled.  In particular, it may be useful to observe nearby SNe
Ia with a range of filters specifically designed to match ``blueshifted'' $I$
or $R$ standard filters for a sample of redshifts (e.g., for $z$ =  0.3, 0.4,
0.6, 0.7;  the current data in $B$ and $V$ may serve for ``blueshifted'' $R$ at
$z \approx 0.5$ and 0.2, or $I$ at $z \approx 0.8$ and 0.5). This would allow
an accurate $K$-correction interpolation table to be constructed.  Note that
this $K$-correction work requires a well-calibrated data set, since any
wavelength-dependent error in the $K$-corrections could mimic
redshift-dependent changes in magnitude, and hence confound the measurements of
$\Omega_M$ and $\Omega_\Lambda$.
 
In practice, actual telescope observing time is, of course, always 
significantly longer than the theoretical predicted time.  These time estimates
are intended to convey the scale of this observing program;
it is an ambitious but practicable program.

\section{Discussion}

\vspace{-0.27in}
As with the other methods for determining the cosmological constant
discussed in the introduction, this approach depends on results from an 
entire research program.  More nearby SNe Ia must be discovered and
studied, as expected from a few projects (e.g. Hamuy et al 1993b;
Muller et al 1992).  This will make it possible to test and refine the
criteria used to distinguish ``normal'' un-extincted SNe Ia, 
to further develop lightcurve
decay-time/shape calibration, and to determine the true
intrinsic magnitude distribution.  Distant SNe Ia must also be discovered 
before maximum light 
on a regular basis (e.g., Perlmutter {\em et al.} 1994a, 1994b), 
and the observational effort
necessary to study them as outlined in this paper will not be trivial.
Both the nearby and distant SNe Ia will contribute to the tests for 
evolution.
Finally, careful photometric and spectral work will still be needed to
ensure that the uncertainty in the $K$-corrections is negligible compared
to the other sources of error. Given
that research programs are already underway in all of these domains, this
approach to the measurement of $\Lambda$ and $\Omega_M$ may soon be feasible.

This work was supported in part by the National Science Foundation
(ADT-88909616), U.S. Dept. of Energy (DE-AC03-76SF000098), and 
Swedish Natural Science Research Council.


\clearpage

%
%


%

\clearpage

%


\begin{thebibliography}{}


\bibitem{BranchMiller}
{ Branch,D. \& Miller,D.L.}
\newblock {1993. Ap.J.Lett.405:L5}.

\bibitem{BranchVandBerg}
{ Branch, D. \& van den Bergh, S.}
\newblock {1993. A.J. 105:2231}.


\bibitem{carroll}
{Carroll,S.M., Press,W.,H. \& Turner,E.L.}
\newblock {1992. Annu.Rev.Astrophys.30:499}.

\bibitem{colgate}
{Colgate, S.A.}
\newblock {1979. Ap.J.232:404-408}.

\bibitem{hamuy}
{Hamuy, M., Phillips, M.M., Wells, L.A., \& Maza, J.}
\newblock{1993. PASP, 105:787}

\bibitem{hamuy2}
{Hamuy, M., et al.}
\newblock{1993b. A.J. 106:2392}

\bibitem{Hamuy3}
{Hamuy, M., Phillips, M.~M., Maza, J., Suntzeff, N.B., Schommer, R.,
\& Aviles, R.}
\newblock {1994. A.J., in press}

\bibitem{Kim}
{ Kim, A., Goobar, A., \& Perlmutter, S.}
\newblock {1995. Lawrence Berkeley Laboratory Report No. LBL-34498; to be
submitted to PASP}.

\bibitem{Kolb}
{ Kolb,E.W \& Turner,M.S.}
\newblock {1990. The Early Universe, Addison-Wesley, Redwood City,CA}.

\bibitem{Massey}
{ Massey, P., Armandroff, T., Harmer, C., Jacoby, G., Schoening, B., \&
Silva, D.}
\newblock {1995. Direct Imaging Manual for Kitt Peak}.

\bibitem{Muller}
{Muller, R.~A.,  Newberg, H.~J.~M., Pennypacker, C.~R., Perlmutter, S., 
Sasseen, T.~P., , and  Smith, C.~K.}
\newblock {1992. Ap.J.Lett, 384:L9--L13}.

\bibitem{Keck}
{Nelson , J.E., Mast, T.S., \& Faber,S.M. }
\newblock{1985. The Design of the Keck Observatory and Telescope,
 Keck Observatory Report No.90}.

\bibitem{Norgaard}
{Norgaard-Nielson, H.~U., Hansen, L., Jorgensen, H.~E., Salamanca, A.~A., 
Ellis, R.~S., \& Couch, W.~J. }
\newblock{1989. Nature, 339:523}.


\bibitem{sn1992bi}
{Perlmutter,S.,Pennypacker,C., Goldhaber, G. Goobar, A. et al.}
\newblock {1994a. Ap.J.Lett. In press}.

\bibitem{sne1994}
{Perlmutter,S., et al.}
\newblock {1994b. International Astronomical Union Circular, nos. 
5956 and 5958}.

\bibitem{phillips}
{Phillips, M.M.}
\newblock {1993. Ap.J.Lett.413:L105}.

\bibitem{riess}
{Riess, A. G., Press, W. H., \& Kirshner, R. P. }
\newblock {1994.  Ap.J., submitted}

\bibitem{sandage}
{Sandage, A. \& Tammann, G.A.}
\newblock {1993. Ap.J.415:1-9}.

\bibitem{Schramm}
{Schramm,D.N.}
\newblock {1990. Astrophysical Ages and Dating Methods: Gif sur Yvette:Edition
  Frontiers}.

\bibitem{sollheim}
{Sollheim, J.E.}
\newblock {1966. MNRAS 133:231}.

\bibitem{stabell}
{Stabell, R. , Refsdal, S. }.
\newblock {1966. MNRAS 132:379}.

\bibitem{refsdal}
{Stabell, R. , Refsdal, S., Lange, F.G. }.
\newblock {1967. MNRAS 71:143}.

\bibitem{vaughan}
{Vaughan, T., Branch, D., Miller, D., \& Perlmutter, S. }
\newblock {1994.  Ap.J., In press}.

\bibitem{vaughan2}
{Vaughan, T., Branch, D., \& Perlmutter, S. }
\newblock {1994.  preprint}.

\end{thebibliography}
\end{document}